\documentstyle[epsfig]{mn}

\begin{document}


\title[Neural networks and component separation in sky maps]
{Neural networks and separation of Cosmic Microwave Background and 
astrophysical signals in sky maps}

\author[Baccigalupi et al.]
{C. Baccigalupi$^1$, L. Bedini$^2$,
C. Burigana$^3$, G. De Zotti$^4$, A. Farusi$^2$,
D. Maino$^5$,
\and
M. Maris$^5$, F. Perrotta$^1$,
E. Salerno$^2$, L. Toffolatti$^{6}$, A. Tonazzini$^2$\\
$^1$SISSA/ISAS, Astrophysics Sector, Via Beirut, 4,
I-34014 Trieste, Italy. Email {\tt bacci@sissa.it,perrotta@sissa.it}\\
$^2$IEI-CNR, Via Alfieri, 1, I-56010 Ghezzano, Pisa, Italy.
Email {\tt $<$name$>$@iei.pi.cnr.it}\\
$^3$ITeSRE-CNR, Via Gobetti, 101, I-40129 Bologna,
Italy. Email {\tt burigana@tesre.bo.cnr.it}\\
$^4$Oss. Astr. Padova, Vicolo dell'Osservatorio 5, 35122 Padova, Italy.
Email {\tt dezotti@pd.astro.it}\\
$^5$Oss. Astr. Trieste, Via G.B. Tiepolo, 11, I-34131
Trieste, Italy. Email {\tt <name>@ts.astro.it}\\
$^{6}$Dpto. de F\'\i{sica}, c. Calvo Sotelo s/n, 33007 Oviedo, Spain
Email {\tt toffol@pinon.ccu.uniovi.es}}

\maketitle

\begin{abstract}
We implement an Independent Component Analysis (ICA) 
algorithm to separate signals of different origin in sky maps 
at several frequencies. Due to its 
self-organizing capability, it works without prior assumptions
either on the frequency dependence or on the angular power spectrum of 
the various signals; rather, it learns directly from the
input data how to identify the statistically independent components, on 
the assumption that all but, at most, one of them have non-Gaussian 
distributions.

We have applied the ICA algorithm to simulated patches of the sky 
at the four frequencies (30, 44, 70 and 100 GHz) of the Low Frequency 
Instrument (LFI) of ESA's {\sc Planck} satellite. Simulations include 
the Cosmic Microwave Background (CMB), the synchrotron and thermal dust 
emissions and extragalactic radio sources. The effects of detectors angular 
response functions and of instrumental noise have been ignored in this 
first exploratory study. The ICA algorithm reconstructs the spatial 
distribution of each component with rms errors of about 1\% for the CMB and  
of about $10\%$ for the, much weaker, Galactic components. 
Radio sources are almost completely recovered down to a flux limit 
corresponding to $\simeq 0.7\sigma_{CMB}$, where 
$\sigma_{CMB}$ is the rms level of CMB fluctuations. 
The signal recovered has equal quality on all scales
larger then the pixel size. In addition, we show that 
for the strongest components (CMB and radio sources) 
the frequency scaling is recevered with percent precision. 
Thus, algorithms of the type presented here appear to be very promising 
tools for component separation. On the other hand, we have been dealing 
here with an highly idealized situation. Work to include instrumental 
noise, the effect of different resolving powers at different frequencies 
and a more complete and realistic characterization of astrophysical 
foregrounds is in progress.
\end{abstract}

\section{Introduction}
\label{introduction}

Maps produced by large area surveys aimed at imaging primordial 
fluctuations of the Cosmic Microwave Background (CMB) contain 
a linear mixture of signals by several astrophysical 
and cosmological sources (Galactic synchrotron, free-free and dust emissions, 
both from compact and diffuse sources, extragalactic sources, 
Sunyaev-Zeldovich effect in clusters of galaxies or by inhomogeneous 
re-ionization, in addition to primary and secondary CMB anisotropies) 
convolved with the spatial and spectral responses of the antenna and of the 
detectors. In order to exploit the unique cosmological information encoded in 
the CMB anisotropy patterns as well as the extremely interesting astrophysical 
information carried by the foregound signals, we need to accurately separate 
the different components. 

A great deal of work has been carried out in recent years in this area 
(see de Oliveira-Costa \& Tegmark 1999, and references therein; 
Tegmark et al. 2000).  
The problem of map denoising has been tackled with the
wavelets analysis on the whole sphere \cite{TENORIO} and on
sky patches \cite{SANZb}. Algorithms to single out the CMB and the 
various foregrounds have been developed \cite{WF,HOBSON,TE}. 
In these works, Wiener filtering (WF) and the maximum entropy method (MEM) 
have been applied to simulated data from the {\sc Planck} satellite,
taking into account the expected performances of the instruments. 
Assuming a perfect knowledge of the frequency dependence 
of all the components, as well as priors for the
statistical properties of their spatial pattern,
these algorithms are able to recover the 
the strongest components, at the best {\sc Planck} resolution. 

We adopt a rather different approach, considering denoising and  
deconvolution of the signals on one side and component separation on the other  
as separate steps in the data analysis process, and focus here on the latter 
step only, presenting a 'blind separation' method, based on 'Independent  
Component Analysis' (ICA) techniques. The method does not require any a priori 
assumption 
on spectral properties and on the spatial distribution of the various 
components, but only that they are statistically independent 
and all but at most one have a non-Gaussian distribution.
It is important to note that this is in fact the physical 
system we have to deal with: surely all the foregrounds are non-Gaussian, 
while the CMB is expected to be a nearly Gaussian fluctuation 
field for most of the candidate theories of the early 
universe. 

The paper is organized as follows. In Section 2 we introduce the relevant 
formalism and briefly review methods applied in previous works.  
In Section 3 we outline the ICA algorithm in a rather general 
framework, since it may be useful for a variety of astrophysical 
applications. In Section 4 we describe our simulated maps. In 
Section 5 we give some details on our analysis and present the results. 
In Section 6 we draw our conclusions and list some
future developments.

\section{Formalism and previous approaches}
\label{formalism}

We assume that the frequency spectrum of radiation components 
(referred to as sources) is independent 
of the position in the sky. Since we deal here with relatively small patches 
of the sky, we adopt Cartesian coordinates, $(\xi , \eta )$. 
The function describing the i-th source then writes
\begin{equation}
\label{sorg}
	\tilde{s}_{i}(\xi , \eta , \nu ) = s_{i}(\xi , \eta )\cdot
	{\cal F}_{i}(\nu ) \hspace{10mm}i = 1, \ldots , N
\end{equation}
where $N$ is the number of independent sources and ${\cal F}_{i}(\nu )$ 
is the emission spectrum.  

The signal received from the point $(\xi,\eta)$ in the sky is
\begin{equation}
\label{freq}
	\tilde{x}(\xi,\eta , \nu ) = \sum_{i=1}^{N}s_{i}(\xi , \eta
)\cdot
	{\cal F}_{i}(\nu )
\end{equation}
Suppose that the instrument has $M$ channels,
with spectral response functions $t_{j}(\nu)$, $j=1,\ldots M$
centered at different frequencies, and that the beam patterns
are independent of frequency within each passband.
Let beam patterns be described by the space-invariant
PSF's $\ h_{j}(\xi,\eta)$, so that the maps are produced by a
linear convolutional mechanism. (Note that this is an additional 
simplifying assumption since in real experiments a position dependent 
defocussing related to the chosen scanning strategy may occur.) 
Then, the map yielded by j$^{th}$ channel is:
$$
x_{j}(\xi,\eta) = \int_{}^{} h_{j}(\xi - x, \eta - y)t_{j}(\nu)\cdot
$$
$$
\cdot\sum_{i=1}^{N}s_{i}(x,y)
	{\cal F}_{i}(\nu) dx dy d\nu +\epsilon_{j}(\xi ,\eta) =
$$
\begin{equation}
	\label{chan}
	= \tilde{x}_{j}(\xi,\eta) * h_{j}(\xi,\eta) +
	\epsilon_{j}(\xi ,\eta)\ ,
\hspace{10mm}j=1,\ldots,M\ ,
\end{equation}
where:
\begin{equation}
	\label{freq2}
	\tilde{x}_{j}(\xi,\eta) = \sum_{i=1}^{N}a_{ji}
	\cdot s_{i}(xi ,\eta )\ ,
	\hspace{10mm}j=1,\ldots,M\ ,
\end{equation}
\begin{equation}
	\label{aentries}
	a_{ji} = \int_{}^{}{\cal F}_{i}(\nu)
	t_{j}(\nu) d\nu \ ,\ \
	j=1,\ldots,M;\ i=1,\ldots,N\ ,
\end{equation}
$*$ denotes linear convolution and
$\epsilon_{j}(\xi ,\eta)$ represents the instrumental noise.
Eq.~(\ref{freq2}) can also be written in matrix form:
\begin{equation}
\label{freqmat}
	\tilde{\textbf{x}} (\xi,\eta) = A \textbf{s}(\xi , \eta )
\end{equation}
where the entries of the $M\times N$ matrix $A$ are given by 
Eq.~(\ref{aentries}).

The unknowns of our problem are the $N$ functions $s_{i}(\xi , \eta )$,
and the data set is made of the $M$ maps $x_{j}(\xi,\eta)$ in
Eq.~(\ref{chan}). Besides the measured data, we also know 
the instrument beam-patterns $h_{j}(\xi,\eta)$, and, more or less
approximately (depending on the specific source), the coefficients
$a_{ji}$ in Eq.~(\ref{freq2}).

Eq.~(\ref{chan}) can be easily rewritten in the Fourier space:
\begin{equation}
\label{fourierdata}
	X_{j}(\omega_{\xi},\omega_{\eta}) 
	=\sum_{i=1}^{N}R_{ji}(\omega_{\xi},\omega_{\eta})
	S_{i}(\omega_{\xi},\omega_{\eta}) +
	{\cal E}_{j}(\omega_{\xi},\omega_{\eta})\ ,
\end{equation}
where the capital letters denote the Fourier transforms of the
corresponding lowercase functions, and
\begin{equation}
R_{ji}(\omega_{\xi},\omega_{\eta})={\cal H}_j(\omega_{\xi},\omega_{\eta})
a_{j i } \ \ ,
\end{equation}
${\cal H}_j $ being the Fourier transform of the beam profile $h_j$. \\
Eq.~(\ref{fourierdata}) can thus be rewritten in matrix form:
\begin{equation}
\label{eqnbase}
	{\bf X} = R {\bf S} + {\bf {\cal E}}\ .
\end{equation}
The above equation must be satisfied by each Fourier mode
$(\omega_{\xi},\omega_{\eta})$, independently.
The aim is to recover the true signals
$S_{i}(\omega_{\xi},\omega_{\eta})$ constituting
the column vector ${\bf S}$. If the matrix $A$ in Eq.~(\ref{freqmat})
is known exactly then, in the absence of noise, the problem 
reduces to a linear inversion of Eq.~(\ref{eqnbase}) for 
each Fourier mode.

In practice, however, ${\cal H}_{j}$ vanishes for some Fourier mode. 
For these modes the entire j-th row of the matrix $R$ also 
vanishes, and $R$ may become a non-full-rank matrix.
An inversion based on statistical approaches built on {\it a priori} 
knowledge is thus needed.

In the following two subsections we briefly describe two such
approaches, and in the third one we briefly recall a technique  
so far mostly exploited for the denoising problem and for extraction 
of extragalactic sources. 

\subsection{The maximum entropy approach}
\label{the}

The Maximum Entropy Method (MEM) for the reconstruction of images is 
based on a Bayesian approach to the problem (Skilling 1988, 1989; 
Gull 1988). Let 
${\bf X}$ be a vector of $M$ observations
whose probability distribution $P({\bf X}|{\bf S})$ depends on the
values of $N$ quantities  ${\bf S}={S_1,...,S_N}$. 

Let us $P({\bf S})$ be the {\it prior}  probability
distribution of ${\bf S}$, telling us what is known about ${\bf S}$
without knowledge of the data. Given the data ${\bf X }$, 
Bayes' theorem states that the
conditional distribution of ${\bf S}$ (the {\it posterior} distribution of
${\bf S}$) is given by the product of the likelihood of the data,
$P({\bf X} | {\bf S})$, with the prior:
\begin{equation}
\label{Bayes}
P( {\bf S }| {\bf X} )= z \cdot P({\bf X}|{\bf S}) P({\bf S})\ ,
\end{equation}
where $z$ is a normalization constant. 

An estimator ${\hat {\bf S}}$ of the true signal vector can be constructed  
by maximizing the posterior probability
$ P( {\bf S }| {\bf X} ) \propto  P({\bf X}|{\bf S}) P({\bf S})$.
However, while the likelihood in Eq.~(\ref{Bayes}) is easily 
determined once the noise and signal covariance matrices are known,
the appropriate choice of the prior distribution for the model considered 
is a major problem in the Bayesian approach:  
since Bayes' theorem  is simply a rule for manipulating probabilities,
it cannot by itself help us to assign them in the first place, so one has
to look elsewhere. The MEM is a consistent
variational method for the assignment of probabilities under certain
types of constraints that must refer to the probability distribution
directly. 

The Maximum Entropy principle states that if one has some
information $I$ on which the probability distribution is based, one can
assign a probability distribution to a proposition $i$ such that $P(i|I)$
contains only the information $I$ that one actually possesses. This
assignment is done by maximizing the Entropy
\begin{equation}
H \equiv - \sum_{i=1}^{N} P(i|I)logP(i|I)
\end{equation}
It can be seen  that when nothing is known except that the probability
distribution should be normalized, the Maximum Entropy principle yields 
the uniform prior. In our case the proposition $i$ represents
{\bf S}, and the information $I$ is the assumption of signal
statistical independence. The standard application of the method 
considered strictly positive signals (Skilling 1988, 1989; 
Gull 1988); the extension to 
the case of CMB temperature fluctuations, which can be both positive and 
negative, was worked out by Hobson et al. (1998).

The construction
of the entropic prior requires, in general, the
knowledge of the frequency dependence of the components
to be recovered as well as of the signal covariance matrix
${\bf C }({\bf k}) =< {\bf S}({\bf k}) {\bf S}^{\dagger}({\bf k})>$, 
with the average taken on all the possible realizations.

\subsection{The multifrequency Wiener filtering}

If a Gaussian prior is adopted, the Bayesian approach gives the 
multifrequency Wiener filtering (WF) solution \cite{WF}. In 
in this case too an estimator of the signal vector is obtained by maximizing 
the posterior probability in Eq.~(\ref{Bayes}), 
given the signal covariance matrix ${\bf C}({\bf k})$.

The Gaussian prior probability distribution for the signal has the form
\begin{equation}
P( {\bf S}) \propto \exp (-{\bf S}^{\dagger} {\bf C}^{-1}{\bf S}) \ \ .
\end{equation}
The estimator ${\bf{\hat{\rm S}}}$ is linearly related to the data vector
${\bf{\hat{\rm X}}}$ through the Wiener matrix
${\bf W} \equiv ( {\bf C}^{-1}+{\bf R}^{\dagger} {\bf N}^{-1}{\bf
R})^{-1}$, where ${\bf R}$ corresponds to the matrix in
(\ref{eqnbase}) and ${\bf N}({\bf k})=<{\bf \epsilon}({\bf k})
{\bf \epsilon}^{\dagger}({\bf k})>$ is the noise covariance
matrix:
\begin{equation}
{\hat {\bf S}} = {\bf W}{\bf X} \ \ .
\label{Wiener}
\end{equation}
The ${\bf W}$ matrix has the role of  a linear filter;
again, its construction requires an {\it a
priori} knowledge  of the spectral behavior of the signals.

This method is endangered by the clear non-Gaussianity of foregrounds. 

\subsection{Wavelet methods }

The development of wavelet techniques for signal processing
has been very fast in the last ten years \cite{JAWERTH}. 
The wavelet approach is conceptually very simple: 
whereas the Fourier transform is highly inefficient in dealing with 
the local behavior, the wavelet transform is able to 
introduce a good space-frequency localization, 
thus providing information on the contributions coming from different
positions and scales. 

\noindent
In one dimension, we can define the {\it analyzing} wavelet as
$\Psi (x; R, b) \equiv R^{-1/2}\psi[(x - b)/R]$, dependent on two
parameters, dilation ($R$) and translation ($b$); 
$\psi (x)$ is a one-dimensional function satisfying the following
conditions: a) $\int_{-\infty}^{\infty}dx\,\psi (x) = 0$, b)
$\int_{-\infty}^{\infty}dx\,{\psi}^2 (x) = 1$ and
c) $\int_{-\infty}^{\infty}dk\,|k|^{-1}\psi^2 (k) < \infty $,
where $\psi (k)$ is the Fourier transform of $\psi (x)$.
The wavelet $\Psi$ operates as a mathematical
microscope of magnification $R^{-1}$ at the space point $b$.
The wavelet coefficients associated to a one-dimensional
function $f(x)$ are:

\begin{equation}
w(R, b) = \int dx\, f(x)\Psi (x; R, b)\; .
\end{equation}

\noindent
The computationally faster algorithms for the wavelet analysis of 
2-dimensional images are those based on
Multiresolution analysis \cite{MALLAT} or on 2D wavelet analysis
\cite{LEMARIE}, using tensor products of one-dimensional wavelets.
The discrete Multiresolution analysis entails the definition of a
one-dimensional {\it scaling} function $\phi$, normalized as
$\int_{-\infty}^{\infty}dx\,\phi (x) = 1$ \cite{OGDEN}.
Scaling functions act as low-pass filters whereas wavelet functions
single out one scale. 
The 2D wavelet method \cite{SANZb} is based on two scales,
providing therefore more information on different resolutions (defined
by the product of the two scales) than the Multiresolution one.

\medskip\noindent
Recently, wavelet techniques
have been introduced in the analysis of CMB data.
Denoising of CMB maps has been performed on patches of the sky of
$12^{\circ}.8\times 12^{\circ}.8$
using either multiresolution techniques \cite{SANZa}
and 2D wavelets \cite{SANZb}, as well as on the whole celestial
sphere \cite{TENORIO}. As a first step, 
maps with the cosmological signal plus a Gaussian instrumental 
noise have been considered. 

Denoising of CMB maps has been carried out by using a 
signal--independent prescription, the SURE thresholding method
\cite{DONOHO}. The results are model independent
and only a good knowledge of the noise affecting
the observed CMB maps is required, whereas nothing has to be assumed
on the nature of the underlying field(s).
Moreover, wavelet techniques are highly efficient in localizing noise 
variations and features in the maps. 

The wavelet method is able to improve the signal-to-noise ratio by a 
factor of 3 to 5; correspondingly, the error on $C_{\ell}$'s 
derived from denoised maps is about 2 times lower than that obtained 
with the WF method.

Wavelets were also successfully applied to the detection of 
point sources in CMB maps in the presence of the cosmological signal 
and of instrumental noise \cite{TENORIO}; 
more recently, successfull results on source
detection have also been obtained in presence of diffuse galactic
foregrounds \cite{CAYON}.
The results are comparable to those obtained 
with the filtering method presented by \cite{TEGMARKCOSTA} which, 
however, relies on the assumption that all the underlying fields are
Gaussian.

\section{The ICA approach}
\label{the ICA}

We present here a rather different approach, characterized by the capability
of working `blindly' i.e. without prior knowledge of spectral 
and spatial properties of the signals to be separated. The method is of 
interest for a broad variety of signal and image processing applications, 
i.e. whenever a number of
source signals are detected by multiple transducers, and the
transmission channels for the sources are unknown, so that each
transducer receives a mixture of the source signals with unknown
scaling coefficients and channel distortion.

In this exploratory study we confine ourselves to the case of  
simple linear combinations of unconvolved source signals 
\cite{AMARI,BELL}. The problem can be stated as follows: a set of $N$
signals is input to an unknown frequency
dependent multiple-input-multiple-output linear instantaneous system,
whose $M$ outputs are our observed signals. We use the term
{\it instantaneous} to denote a system whose output at a given point
only depends on the input signals at the same point.
Our objective is to find a stable {\it reconstruction system} to estimate 
the original input signals with no
prior assumptions either on the signal distributions
or on their frequency scalings. The problem in its general form is
normally unsolvable, and a ``working hypothesis" must be made. The 
hypothesis we make is the mutual {\it statistical independence} 
of our source signals, whatever their actual distributions are.
Several solutions have been proposed for this problem, each based on
more or less sound principles, not all of which are typical of
classical signal processing. Indeed, information theory, neural
networks, statistics and probability have played an important part in
the development of these techniques.


We do not consider here specific instrumental
features like beam convolution and noise contamination,
leaving the specialization of the ICA method to 
specific experiments to future work; this allows us to
highlight the capabilities of this approach, able to work
in conditions where other algorithms would not be viable.
Therefore, we adopt Eq.~(\ref{freqmat}) as our data model, just dropping the
tilde accent on vector ${\bf x}$. Also, the instrumental noise term in 
Eq.~(\ref{fourierdata}) will be neglected. 

It can be proved that, to solve the problem described above, the
following hypotheses should be verified \cite{AMARI,COMON}:
\begin{itemize}
\item All the source signals are statistically independent;
\item At most one of them has a Gaussian distribution;
\item $M \geq N$;
\item Low noise.
\end{itemize}

The last two assumptions can be somewhat relaxed by choosing suitable
separation strategies. As far as independence is concerned, roughly
speaking, we may say that the search for an ICA model from non-ICA data (i.e.
data not coming from independent sources) should give the most
`interesting' (namely, the most structured) projections of the data
\cite{HYVARINEN,FRIEDMAN}.
This is not equivalent to say that separation is achieved; however, we
have seen from our experiments that a good separation can be obtained
even for sources that are not totally independent. The second
assumption above tells us that Gaussian sources cannot be separated.
More specifically, they can only be separated up to an orthogonal
transformation. In fact, it can be shown that the joint probability of 
a mixture of Gaussian signals is invariant to orthogonal 
transformations. This means that if independent components are found
from Gaussian mixtures, then any orthogonal transformation of them 
gives mutually independent components.

Many strategies have been adopted to solve the separation problem on
the basis of the above hypotheses, all based on looking for a
set of independent signals, which can be shown to be the original
sources. A formal criterion to test independence, from which all the
separating strategies can be derived, is described later in this section.

In order to recover the original source signals from the observed
mixtures, we use a separating scheme in the form of a feed-forward
neural network. The observed signals are 
input to an $N\times M$  matrix $W$, referred to as the {\it the synaptic
weight
matrix}, whose adjustable entries, $w_{ij}, i=1, \ldots N, j=1,
\ldots M$, are updated
for every sample of the input vector $\textbf{x}(\xi , \eta )$ (at
step $\tau$)
following a suitable {\it learning algorithm}.
The output of matrix $W$ at step $\tau$ will be:
\begin{equation}
\label{weight}
	\textbf{u}(\xi ,\eta ,\tau ) = W(\tau )
	\textbf{x}(\xi ,\eta )\ .
\end{equation}
$W(\tau )$ is expected to converge
to a true separating matrix, that is, a matrix whose output is a {\it
copy} of the inputs, for every point $(\xi , \eta )$. Ideally,
this final matrix $W$ should be such that $WA = I$,
where $I$ is the $N\times N$
identity. As an example, if $M = N$, we should have $W = A^{-1}$.
There are, however, two basic indeterminacies in our problem:
ordering and scaling. Even if we are able to extract $N$ independent
sources from $M$ linear mixtures, we cannot know {\it a
priori} the order in which they will be arranged, since this corresponds
to unobservable permutations of the columns of matrix $A$. Moreover,
the scales of the extracted signals are unknown, because when a
signal is multiplied by some scalar constant, the effect is the same
as of multiplying by the same constant the corresponding column of the
mixing matrix. This means that $W(\tau )$ will converge, at best, to a
matrix $W$ such that:
\begin{equation}
\label{converg}
	WA = PD\ ,
\end{equation}
where $P$ is any $N\times N$ permutation matrix, and $D$ is a
nonsingular diagonal scaling
matrix. From Eqs.~(\ref{freqmat}), (\ref{weight}) and
(\ref{converg}) we thus have:
\begin{equation}
\label{invert}
	{\bf u} = W{\bf x} = WA{\bf s} = PD{\bf s}\ .
\end{equation}
That is, as anticipated, each component of $\bf u$ is a scaled version
of a component of $\bf s$, not necessarily in the same order.
This is not a serious inconvenience in our application, since we should
be able to recover the proper scales for the separated sources from
other pieces of information, for example matching with independent
lower resolution observations like those of COBE on the 
case of  MAP and {\sc Planck}.
If $A$ was known, the performance of the separation algorithm could
be evaluated by means of the 
matrix $WA$. If the separation is perfect, this matrix has only one
nonzero element for each row and each column.
In any non-ideal situation each row and column of $WA$ should contain only 
one dominant element.

In all the cases treated
here we assume $M \geq N$, but we consider
the case where $N$, although smaller than $M$, is not known.

The mutual statistical independence of the source signals can be
expressed in terms of a separable joint probability density function
$q({\bf s})$:
\begin{equation}
\label{separable}
	q({\bf s}) = \prod_{j=1}^{N}q_{j}(s_{j})
\end{equation}
where $q_{j}(s_{j})$ is the marginal probability density of the
$j^{th}$ source.

Various algorithms can be used to learn the 
matrix $W$. All these algorithms can be derived from a unified principle
based on the Kullback--Leibler (KL) divergence between the joint
probability
density of the output vector $\bf u$, $p_{U}(\bf u)$, and a
function $q({\bf u})$, which should
be suitably chosen among the ones of the type of Eq.~(\ref{separable}).
The KL divergence between the two functions mentioned
above may be written as a function of the matrix $W$, and can be
considered as a cost function in the sense of Bayesian statistics:
\begin{equation}
\label{kullei}
	R(W) = \int_{}^{}p_{U}({\bf u}) {\rm log} \frac{p_{U}({\bf
u})}{q({\bf u})} d {\bf u}\ .
\end{equation}
It can be proved that, under mild
conditions on $q({\bf u})$, $R(W)$ has a global minimum
where $W$ is such that $WA = PD$.
The different possible choices for $q({\bf s})$ lead to the
different particular learning strategies proposed in the
literature \cite{AMARI,AMARI2,BELL}.

The {\it uniform gradient} search
method, which is a gradient-type algorithm, takes into account
the Riemannian metric structure of our objective parameter space,
which is the set of all nonsingular matrices $W$ \cite{AMARI}.
In a general case, where the number $N$ of sources is only known to
be smaller than the number of observations, the following formula is derived:
$$
	W(\tau +1)=W(\tau ) +
$$
\begin{equation}
\label{alg2}
+\alpha(\tau )\cdot[\Lambda - {\bf u}(\tau ){\bf u}^{T}(\tau )
	-{\bf f}({\bf u}(\tau )) {\bf u}^{T}(\tau )]W(\tau )\ ,
\end{equation}
where $\Lambda$ is a $M\times M$ diagonal matrix:
\begin{equation}
	\Lambda ={\rm diag} { [ (u_{1} + f_{1}(u_{1}))u_{1}]
	\ldots [(u_{M} + f_{M}(u_{M}))u_{M} ] }\ .
\end{equation}
Pixel by pixel,
the $M\times M$ matrix $W$ is multiplied by the M--vector {\bf x}, and
gives vector ${\bf u}$ as its output. This output is transformed through
the nonlinear vector function ${\bf f}({\bf u})$, and the result is
combined with $\bf u$ itself to build the update to matrix $W$,
through Eq.~(\ref{alg2}). The process has to be iterated by
reading the data maps several times. If $N$ is strictly smaller than
$M$, then $M-N$ outputs can be shown to rapidly converge to zero, or
to pure noise functions.

The convergence properties of this iterative formula are shown to be
independent
of the particular matrix $A$, so that, even a strongly ill-conditioned
system does not affect the convergence of the learning algorithm. In
other words, even when the contributions from some components are very small,
there is no problem to recover them. This property is
called the {\it equivariant property} since the asymptotic properties
of the algorithm are independent of the mixing matrix.
The $\tau$-dependent parameter $\alpha$
is the {\it learning rate}; its value is normally
decreased during the iteration. As far as the choice of $\alpha
(\tau)$ is concerned, a strategy to
learn it and its annealing scheme
is given in Amari et al. (1998);
we have chosen $\alpha (\tau )$ decreasing from $10^{-3}$ to
$10^{-4}$ linearly with the number of iterations.

The final problem is how to choose the function ${\bf f}({\bf u})$. If we
know the true source distributions
$q_{j}(u_{j})$, the best choice is to make $f'_{j}(u_{j}) =
q_{j}(u_{j})$,
since this gives the maximum likelihood estimator. However, the point
is that when $q_{j}(u_{j})$ are specified incorrectly, the algorithm
gives the correct answer under certain conditions. In any case, the
choice of ${\bf f}({\bf u})$ should be made to ensure the existence
of an equilibrium point for the cost function and the stability of the
optimization algorithm. These requirements can be satisfied even
though the nonlinearities chosen are not optimal. A suboptimal choice
for sub-Gaussian source signals (negative kurtosis), is:
\begin{equation}
\label{subgauss}
	f_{i}(u_{i}) = \beta u_{i} + u_{i}|u_{i}|^{2}\ ,
\end{equation}
and, for super-Gaussian source signals (positive kurtosis):
\begin{equation}
\label{supergauss}
f_{i}(u_{i}) = \beta u_{i} + {\rm tanh}(\gamma u_{i})\ ,
\end{equation}
where $\beta \geq 0$ and $\gamma \geq 2$; if one source is 
Gaussian, the above choices remain viable as well. 
In our case, we verified that all the source functions 
except CMB are super-Gaussian,
and thus we implemented the learning algorithm following Eq.(\ref{alg2}),
with the nonlinearities in Eq.~(\ref{supergauss}), and $\beta =0$,
$\gamma = 2$. As already stated, the mean of
the input signal at each frequency is subtracted.
In previous works \cite{AMARI2} the initial matrix was chosen as
$W\propto I$; in that analysis, the image data consisted of a set of
components with nearly the same amplitude.
The initial guess for $W$ affects the computation
time, as well as the scaling of the reconstructed signals
and their order.
Interestingly, we found that adjusting the diagonal elements so that 
they roughly reflect the different weights of the
components in the mixture can speed-up the convergence. For the
problem at hand, the results shown in \S~\ref{blind} have
been obtained starting from $W=$diag[1,3,30,10], for the case of
a $4\times 4 \ W$-matrix, and using only 20 learning steps:
the time needed was about 1 minute on a UltraSparc machine,
equipped with an 300 MHz UltraSparc processor, 256 MBytes RAM,
running down SUN Solaris 7 Operating System, compiling the
{\tt FORTRAN 90} code using SUN Fortran Workshop 5.0

\section{Simulated maps}
\label{database}

We produced simulated maps of the antenna temperature distribution 
with 3'.5 pixel size of a $15^{\circ}\times 
15^{\circ}$ region centered at $l=90^{\circ}$, $b=45^{\circ}$, 
at the four central frequencies of  
the {\sc Planck}/LFI channels \cite{MANDOLESI},
namely 30, 44, 70 and 100 GHz (Fig.~\ref{icasissain}).
The HEALPix pixelization scheme \cite{Heal} was adopted. 
The maps include CMB anisotropies, Galactic synchrotron and dust emissions, 
and extragalactic radio sources.
  
CMB fluctuations correspond to a flat Cold Dark Matter (CDM) model 
($\Omega_{CDM}=.95$, $\Omega_{b}=.05$, three massless neutrino species),  
normalized to the COBE data \cite{SZ}. As it is well known, the CMB
spectrum, in terms of antenna temperature, writes:
\begin{equation}
\label{nucmb}
s^{antenna}_{CMB}(\xi ,\eta ,\nu)=s^{thermod.}_{CMB}(\xi ,\eta ) \cdot
{ \tilde{\nu}^2 e^{\tilde{\nu}} \over (e^{\tilde{\nu}} -1 )^2 }\ ,
\end{equation}
where $\tilde{\nu}={\nu/ 56.8}$ and $\nu$ is the frequency in GHz;
$s^{thermod.}_{CMB}(\xi ,\eta )$ is frequency independent \cite{FIXSEN}. 

As for Galactic synchrotron emission, we have extrapolated the 
408 MHz map with about 1 degree resolution \cite{H}, assuming 
a power law spectrum, in terms of antenna temperature:
\begin{equation}
{\cal F}_{syn}  \propto  \tilde{\nu}^{-n_s}\ ,
\end{equation}
with spectral index $n_s=2.9$.

The dust emission maps with about 6' resolution constructed by Schlegel et 
al. (1998) combining IRAS and DIRBE data have been used as templates for 
Galactic dust emission. The extrapolation to {\sc Planck}/LFI frequencies 
was done assuming a grey-body spectrum:
\begin{equation}
{\cal F}_{dust} \propto {\tilde{\nu}^{m+1} \over
e^{\tilde{\nu}} -1}\ ,
\end{equation}
with $m=2$, $\tilde{\nu}=h \nu / kT_{dust}$, $T_{dust}$ being the dust 
temperature. Although, in general, $T_{dust}$ varies across the sky, 
it turns out to be approximately constant at about $18\,$K in the region 
considered here; we have therefore adopted this value in the above equation.
 
Because of the lack of a suitable template, we have ignored here free-free 
emission, which may be important particularly at 70 and 100 GHz. This 
component needs to be included in future work.

The model by Toffolatti et al. (1998) was adopted for extragalactic radio 
sources, assumed to have a Poisson distribution. An antenna temperaure 
spectral index $n_{\rm rs}=1.9$ was adopted 
$({\cal F}_{\rm rs} \propto  \tilde{\nu}^{-n_{\rm rs}})$.

\begin{table*}
\caption{Input and output frequency scalings of the various components.}
\begin{tabular}{r c c c c c c c c }
\hline
\multicolumn{1}{c}{Frequency}  & \multicolumn{2}{c}{Radio sources} & 
\multicolumn{2}{c}{CMB} & \multicolumn{2}{c}{synchrotron} & 
\multicolumn{2}{c}{dust} \\
\multicolumn{1}{c}{(GHz)} & input & output\qquad & input & output\qquad & 
input & output\qquad & input & output \\ 
\hline
100\quad\quad & 1.00 & 1.00 & 1.00 & 1.00 & 1.00 & 1.00 & 1.00 & 1.00\\
70\quad\quad &  1.97 & 1.95 & 1.14 & 1.14 & 2.81 & 1.36 & 0.68 & 0.93\\
44\quad\quad &  4.76 & 4.70 & 1.22 & 1.23 & 10.8 & 1.72 & 0.35 & 1.93\\
30\quad\quad &  9.86 & 9.70 & 1.26 & 1.26 & 32.8 & -12. & 0.19 & 3.77\\
         &              &      &      &      &      &      &
\label{frequencyin}
\end{tabular}
\end{table*}

\section{Blind analysis and results}
\label{blind}

As it is well known, the strongest signals at the {\sc Planck}/LFI 
frequencies come from the CMB and from radio sources
(although the latter show up essentially as a few high peaks),
whereas synchrotron emission and thermal dust are roughly
1 or 2 orders of magnitude lower, depending on frequency.
Thus we are testing the performances of the ICA algorithm 
with four signals exhibiting very different
spatial patterns, frequency dependences and amplitudes.

Since we are interested in the fluctuation pattern, the mean of the total 
signal (sum of the four components) is set to zero at each frequency. 
We adopt a ``blind'' approach: no information on either the spatial 
distribution or the frequency dependence of the signals is provided to the 
algorithm.

The reconstructed maps of the the four components are shown in 
Fig.~\ref{icasissaout}. Several interesting features may be noticed.
The order of the plotted maps is permuted with respect to the input maps in 
Fig.~\ref{icasissain}, reflecting the order of the ICA outputs:
the first output is synchrotron, the second represents radio
sources, the third is CMB and the fourth is dust. 
All the output maps look  very similar to the true ones; even
synchrotron lower resolution pixels have been reproduced.
In Figs.~\ref{icacmb}, \ref{icadust}, \ref{icasyn} and
\ref{icaradio} we analyze the goodness of the separation by
comparing power spectra and showing scatter plots between
the inputs and the outputs.

\subsection{Signal reconstruction}

For each map, we have computed the angular power spectrum,
defined by the expansion coefficients $C_{\ell}$ of the two
point correlation function in Legendre polynomials.
As is well known, it can conveniently be expressed
in terms of the coefficients of the expansion of the signal
$S$ into spherical harmonics, $S (\theta ,\phi )=
\sum_{\ell m}a_{\ell m}Y_{\ell m}(\theta ,\phi )$:
\begin{equation}
\label{cl}
C_{\ell}={1\over 2\ell +1}\sum_{m}|a_{\ell m}|^{2}\ .
\end{equation}
Such coefficients are useful because
from elementary properties of the Legendre polynomials
it can be seen that the coefficient $C_{\ell}$ quantifies the
amount of perturbation on the angular scale $\theta$ given by
$\theta\simeq 180/ \ell$ degrees. 

The panels on the top of Figs.~\ref{icacmb}, \ref{icasyn},
\ref{icadust}, \ref{icaradio} show the power spectra 
of the input (left) and output (right) signals. The CMB exhibits the 
characteristic peaks on sub-degree angular scales due to acoustic
oscillations of the photon-baryon fluid at decoupling;
the dashed line represents the theoretical model from which
the map was generated, while the solid line is the power
spectrum of our simulated patch: the difference between
the two curves is due to the sample variance corresponding to the CMB Gaussian
statistics. Radio sources are completely different, having all the
power on small scales with the typical shot noise spatial
pattern; dust and synchrotron emissions have power decreasing
on small scales roughly as a power law, as expected \cite{MANDOLESI,PUGET}.
The left-hand side panels on the bottom show the quality factor, defined
as the ratio between true and reconstructed power spectrum
coefficients, for each multipole $\ell$. Due to the limited size of the 
analyzed region, the power spectrum can be defined on  scales
below roughly $2^\circ$. The bottom right-hand side panels are scatter plots 
of the ICA results: for each pixel of the maps, we plotted the value of the
reconstructed image vs. the corresponding input value.  

The reconstructed signals have zero mean and are in unit of the
constant $d$ multiplying each output map, produced during the
separation phase, as mentioned in \S$\,3$: the scale of each
signal is unreproducible for a blind separation algorithm
like ICA. Nevertheless, a lot of information is encoded
into the spatial pattern of each signals, and ultimately
its overall normalization could be recovered exploiting data 
from other experiments.
Therefore, the relation between each true signal and
its reconstruction is
\begin{equation}
\label{relation}
s_{i}^{in}=d\cdot s_{i}^{out}+b\ \ ,\ \ i=1,...,N_{pixels}\ ,
\end{equation}
where $b$ represents merely the mean of the input signal,
that is zero for the CMB and some positive value for the foregrounds. 

To quantify the quality of the reconstruction, we have recovered $d$ and $b$ by
performing a linear fit of output to 
input maps ({\bf s}$^{in}$,{\bf s}$^{out}$) for each signal:
\begin{equation}
\label{alpha}
d={\sum_{i}s^{in}_{i}s^{out}_{i}-
\bar{s}^{in}\cdot\sum_{i}s_{i}^{out}\over
\sum_{i}(s_{i}^{out})^{2}-\bar{s}^{out}\cdot\sum_{i}s_{i}^{out}}
\ \ ,\ \ b=\bar{s}^{in}-d\cdot\bar{s}^{out}\ ,
\end{equation}
where the sums run over all the pixels, and the bar indicates the
average value over the patch; the values of $d$ and $b$, as well
as the linear fits (dashed lines), are indicated
for all the signals in the scatter plot panels.
Also, in the same panels we show the standard deviation of the
fit, that is
\begin{equation}
\label{sigma}
\sigma =\left[{1\over N_{pixels}}
\sum_{i}(s^{in}_{i}-d\cdot s^{out}_{i}-b)^{2}\right]^{1/2}\ .
\end{equation}
A comparison of such quantity with the input signals
(bottom right-hand side panels) gives an estimate  of
the goodness of the reconstruction.  CMB and radio sources are
recovered with percent and $0.1\%$ precision, respectively, while 
the accuracy drops roughly to $10\%$ for the (much weaker) Galactic 
components, synchrotron and dust. Also, the latter 
appear to be slightly mixed; this is likely due to the fact that 
they are somewhat correlated  
so that the hypothesis of statistical independence
is not properly satisfied.

We have also tested to what extent the counts of radio sources are recovered. 
This was done in terms of the relative flux 
\begin{equation}
\label{fracflux}
\Delta s=s/s_{\rm max}\ ,
\end{equation}
$s_{\rm max}$ being the flux of the brightest source. 

In Fig.~\ref{numbercounts} we show the cumulative number of input (dashed) 
and output (solid line) pixels exceeding a given value of $\Delta s$.
The algorithm correctly recovers essentially all sources
with $\Delta s\ge 2\times 10^{-2}$, corresponding to a 
signal of $T_s\simeq 50\,\mu$K, or to a flux density  
$S=(2k_{B}T_s/\lambda^{2})\Delta\Omega\simeq 15\,$mJy, where 
$k_{B}$ the Boltzmann constant,
$\lambda$ the wavelength and $\Delta\Omega$ the solid angle
covered by a pixel, that is $3.5'\times 3.5'\simeq 10^{-6}\,$sr. 
At fainter fluxes the counts are overestimated;
this is probably due to the contamination from the other signals. 
In any case, the flux limit for source detection is surprisingly
low, even lower of the rms CMB fluctuations ($\sigma_{CMB}\simeq 70\mu$K 
at the resolution limit of our maps), 
substantially lower or at least comparable to that
achieved with other methods which require stronger assumptions 
\cite{CAYON,HOBSONII}. 
This high efficiency in detecting point sources illustrates
the ability of the method in taking the maximum advantage
of the differences in frequency and spatial properties of the various 
components. 

On the other hand, we stress that our approach is idealized
in a number of aspects: beam convolution and 
instrumental noise have not been taken into account, 
and the same frequency scaling has been
assumed for all radio sources. 
Therefore more detailed investigations 
are needed to estimate a realistic source detection limit.

Finally note that the quality of the
separation is similar on all scales, as shown by the bottom left-hand side 
panels of Figs.~\ref{icacmb}, \ref{icasyn}, \ref{icadust}, \ref{icaradio}. 
The exception are radio sources, whose true power spectrum goes to zero at 
low $\ell$'s more rapidly than the reconstructed one. 

\subsection{Reconstruction of the frequency dependence}

Another asset of this technique is the possibility of recovering 
the frequency dependence of individual components. 
The outputs can be written as ${\bf u}=W{\bf x}$, 
where ${\bf x}=A{\bf s}$. As previously mentioned, in the ideal case
$WA$ would be a diagonal matrix containing
the constants $d$ for all the signals, multiplied by a permutation
matrix. It can be easily seen that, if this is true, the frequency scalings 
of all the components can be obtained by inverting
the matrix $W$ and performing the ratio, column by column, of
each element with the one corresponding to the row corresponding to 
a given frequency. However, as pointed out in \S$\,3$, 
if some signals are much smaller than others the above 
reasoning is only approximately valid.  This is precisely what is 
happening in our case: we are able
to accurately recover the frequency scaling of the strongest signals,
CMB and radio sources, while the others are lost (see  
Table \ref{frequencyin}).

\section{Concluding remarks and future developments}

We have developed a neural network suitable to implement the 
Independent Component Analysis technique for separating different 
emission components in maps of the sky at microwave wavelengths. 
The algorithm was applied to simulated maps of a 
$15^{\circ}\times 15^{\circ}$ region of sky at 30, 44, 70, 100 GHz,
corresponding to the frequency channels of {\sc Planck}'s 
Low Frequency Instrument (LFI).

Simulations include the Cosmic Microwave Background, 
extragalactic radio sources and Galactic synchrotron and thermal dust 
emission. The various components have markedly
different angular patterns, frequency dependences and amplitudes. 

The technique exploits the statistical independence of the different signals  
to recover each individual component with no prior assumption either on 
their spatial pattern or on their frequency dependence. 
The great virtue of this approach is the
capability of the algorithm to {\it learn} how to recover
the independent components in the input maps. 
The price of the lack of {\it a priori} information
is that each signal can be recovered
multiplied by an unknown constant produced during the
learning process itself. However this is not a substantial
limitation, since a lot of physics is encoded in the spatial
patterns of the signals, and ultimately
the right normalization of each 
component can be obtained by resorting to independent observations.

The results are very promising. The CMB map is recovered with an accuracy 
at the 1\% level. The algorithm is remarkably efficient also in the
detection of extragalactic radio sources: almost all 
sources brighter then 15 mJy at 100 GHz (corresponding to $\simeq 0.7
\sigma_{CMB}$, $\sigma_{CMB}$ being the rms level of CMB fluctuations on 
the pixel scale) are recovered; on the other hand, it must be stressed 
that is not directly indicative of what can be achieved in the analysis 
of Planck/LFI data because the adopted resolution ($3'.5\times 3'.5$) is 
much better than that of the real experiment, 
instrumental noise has been neglected and the 
same spectral slope was assumed for all sources. 

Also the frequency dependences of the strongest components are 
correctly recovered (error on the spectral index of 1\% for the CMB 
and extragalactic sources).

Maps of subdominant signals (Galactic synchrotron and dust emissions) 
are recovered with rms errors of about 10\%; their spectral properties 
cannot be retrieved by our technique. 

The reconstruction has equal quality on
all the scales of the input maps, down to the pixel size.

All this indicates that this technique is suitable for a variety of 
astrophysical applications, i.e. whenever we want to separate 
independent signals from different astrophysical processes occurring 
along the line of sight.

Of course, much work has to be done to better explore the potential 
of the ICA technique. It has to be tested under more realistic 
assumptions, 
taking into account instrumental noise and the effect 
of angular response functions as well as including a more 
complete and accurate characterization of foregrounds.

In particular, the assumption that the spectral properties of each 
foreground component is independent of position will have to be relaxed 
to allow for spectral variations across the sky. Also, it will be 
necessary to deal with the fact that Galactic emissions are correlated. 

The technique is flexible enough to offer good prospects in this respect. 
In the learning stage, the ICA algorithm makes use
of non-linear functions that, case by case, are chosen to
minimize the mutual information between the outputs; 
improvements could be obtained by specializing
the ICA inner non-linearities to our specific needs.
Also, it is possible to take properly into account 
our prior knowledge on some of the signals to recover, still 
taking advantage as far as possible of the ability of 
this neural network approach to carry out a ``blind" separation.
Work in this direction is in progress.

\vskip .1in
We warmly thank Luigi Danese for original suggestions.
We also thank Krzysztof M. G\'orski and all the people who
collaborated to build the HEALPix pixelization scheme
extensively used in this work. Work supported in part by ASI and 
MURST. LT acknowledges financial support from the Spanish DGES,
projects ESP98--1545--E and PB98--0531--C02--01.

\newpage

\onecolumn

\begin{figure}
\vskip 3.in
\caption{Inputs maps used in the ICA
separation algorithm: from top left, in a clockwise sense, 
simulations of CMB, synchrotron, radio sources and dust
emission are shown. Radio sources and dust grey scale are is 
non-linear to better show the signal features. 
}
\label{icasissain}
\end{figure}
\begin{figure}
\vskip 3.in
\caption{Reconstructed maps produced by the
ICA method; the initial ordering has not been conserved in the outputs.
From top left, in a clockwise sense, we can recognize synchrotron, radio
sources, dust and CMB. Radio sources and dust grey scale is 
non-linear as in Fig.1.}
\label{icasissaout}
\end{figure}
\begin{figure}
\caption{Top left: input angular power spectra,
simulated (solid line) and theoretical (dashed line, see text). Top
right: the angular power spectrum of the reconstructed CMB patch. Bottom
left: quality factor relative to the input/output angular spectra.
Bottom right: scatter plot and linear fit (dashed line)
for the CMB input/output maps.
}
\label{icacmb}
\end{figure}
\begin{figure}
\caption{Top panels: angular power spectra for the
simulated input (left) and reconstructed (right) synchrotron map.
Bottom left:  quality factor relative to the input/output angular spectra.
Bottom right: scatter plot and linear fit (dashed line)
for the synchrotron input/output maps.
}
\label{icasyn}
\end{figure}
\begin{figure}
\caption{Top panels: angular power spectra for the simulated input 
(left) and reconstructed (right) dust emission map.
Bottom left:  quality factor relative to the input/output angular spectra.
Bottom right: scatter plot and linear fit (dashed line) for
the dust input/output maps.
}
\label{icadust}
\end{figure}
\begin{figure}
\caption{Top panels: angular power spectra for the
simulated  (left) and reconstructed (right) radio source map.
Bottom left: quality factor relative to the input/output angular spectra.
Bottom right: scatter plot and linear fit (dashed line)
for the radio source emission input/output maps.}
\label{icaradio}
\end{figure}
\begin{figure}
\caption{Cumulative number of pixels as a function of the
threshold $\Delta s$ (see text for more details):
input (dashed line) versus output (solid line).}
\label{numbercounts}
\end{figure}

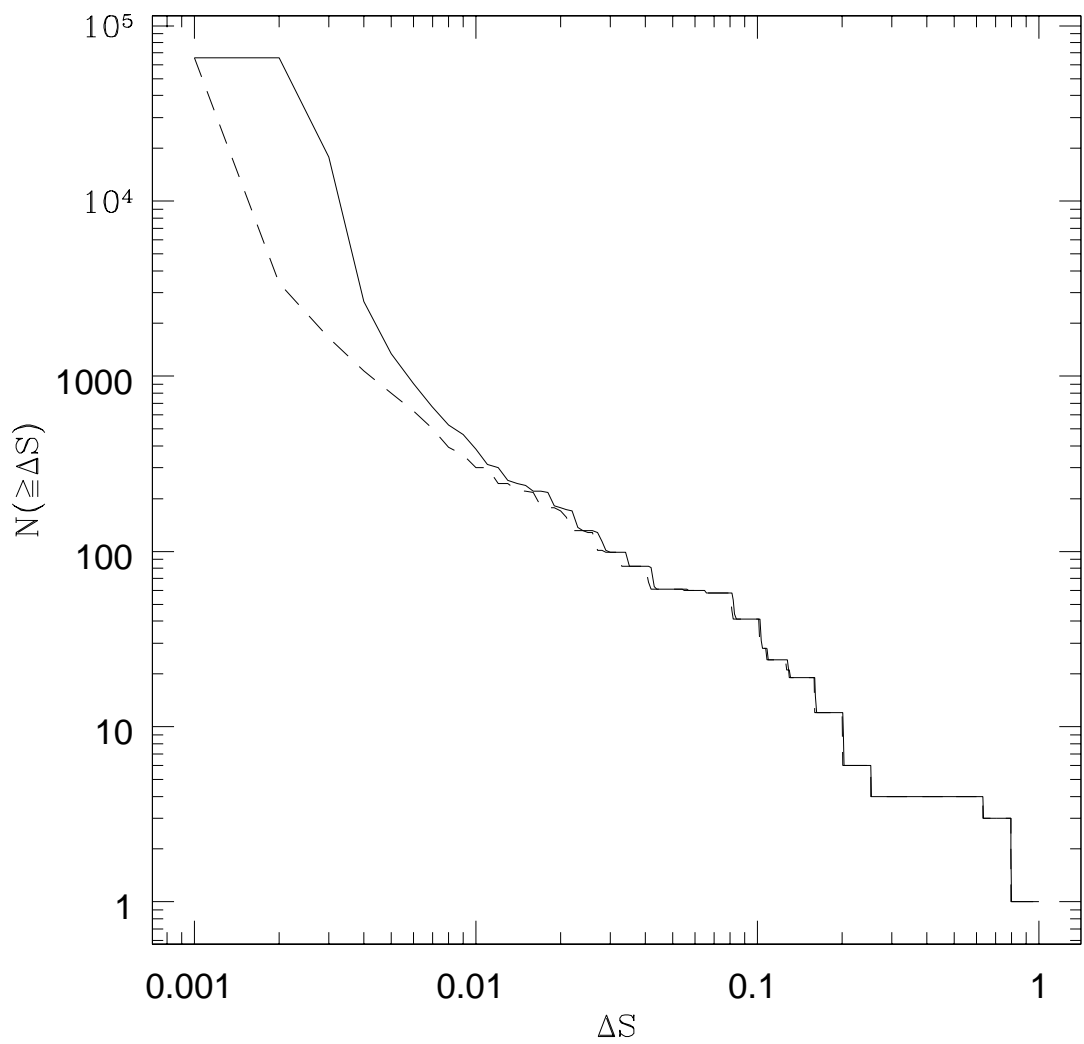
\end{document}